\documentclass[aps,prl,twocolumn,showpacs,floatfix,nofootinbib]{revtex4-1}
\usepackage{epsfig}
\usepackage{color}
\usepackage{enumerate}
\begin{document}

\title{Signatures of lower scale gauge coupling unification in the Standard
Model\\
due to extended Higgs sector}

\author{M.V. Chizhov$^{1,2}$, V.A. Bednyakov$^1$}
\affiliation{$^{\it 1}$Dzhelepov Laboratory of Nuclear Problems,\\
\mbox{Joint Institute for Nuclear Research, 141980, Dubna,
Russia}\\
$^{\it 2}$Centre for Space Research and Technologies, Faculty of
Physics, Sofia University, 1164 Sofia, Bulgaria}

\begin{abstract}
The gauge coupling unification can be achieved at a unification
scale around $5\times 10^{13}$~GeV if the Standard Model scalar
sector is extended with extra Higgs-like doublets. The relevant new
scalar degrees of freedom in the form of chiral $Z^\star$ and
$W^\star$ vector bosons might "be visible"\/ already at about 700
GeV. Their eventual preferred coupling to the heavy quarks explains
the non observation of these bosons in the first LHC run and
provides promising expectation for the second LHC run.
\end{abstract}

 \maketitle

\section{Introduction}

At present the Standard Model (SM) successfully describes all
experimental data in particle physics. Moreover, it is theoretically
consistent and applicable up to the Plank scale,  $M_{\rm Pl}\approx
1.2\times 10^{19}$~GeV. On the other hand, there are many {\it
natural} questions, which cannot be answered within the SM
framework. For example, chiral anomalies are canceled only when
quarks and leptons are considered simultaneously. At the same time
these two sectors are completely independent within the SM.

The hope that the different gauge coupling constants of the
SU(3)$_C\times$SU(2)$_W\times$U(1)$_Y$ SM group meet at a single
unification point has failed~\cite{Amaldi1991}. Therefore, if we,
nevertheless, expect such unification, new physics should be
introduced at some scale above the electroweak unification. In this
paper we will consider the one-loop approximation to the gauge
coupling evolution modifying only the Higgs sector of the SM.

The matter sector of the SM consists of electroweak doublets:
fermionic and bosonic ones. The SM contains only one bosonic doublet
of the Higgs fields. We will assume that the number of the bosonic
doublets $N$ above some scale could be greater than one, while the
number of the fermionic doublets is not modified. At present there
are practically no limitations on the number $N$ of the Higgs
doublets from the precision low-energy measurements~\cite{chile}.

The evolution of the gauge coupling constants in the one-loop
approximation reads
\begin{equation}\label{evolution}
    \alpha_i^{-1}(\mu)=\alpha_i^{-1}(\mu_0)
    -\frac{b_i}{2\pi}\ln\frac{\mu}{\mu_0},
\end{equation}
where the constants $b_i$ are given by the known formulas
\begin{eqnarray}
\label{bi} \nonumber
  b_1 &=& 4+\frac{N}{10}, \\
  b_2 &=& -2\frac{11}{3}+4+\frac{N}{6}, \\
\nonumber
  b_3 &=& -3\frac{11}{3}+4.
\end{eqnarray}

We will start the evolution of $\alpha_i^{-1}$ at the initial point
$\mu_0=M_Z=91.1876\pm 0.0021$~GeV using the most precise physical
constants~\cite{PDG},
\begin{eqnarray}
\nonumber
  \hat{\alpha}(M_Z) &=& 1/127.940\pm 0.014, \\
  \sin^2\hat{\theta}(M_Z) &=& 0.23126\pm 0.00005, \\
\nonumber
  \alpha_s(M_Z) &=& 0.1185\pm 0.0006.
\end{eqnarray}
Then the initial corresponding gauge coupling constants can be
expressed as
\begin{eqnarray}
\nonumber
  \alpha_1^{-1}(M_Z) &=& \frac{3}{5}
  \frac{\cos^2\hat{\theta}(M_Z)}{\hat{\alpha}(M_Z)}=
  59.012\pm 0.014, \\
  \alpha_2^{-1}(M_Z) &=& \frac{\sin^2\hat{\theta}(M_Z)}{\hat{\alpha}(M_Z)}=
  29.587\pm 0.007, \\
\nonumber
  \alpha_3^{-1}(M_Z) &=& \alpha_s^{-1}(M_Z)=8.439\pm 0.043,
\end{eqnarray}
where the factor of 3/5 in the definition of $\alpha_1$ is included
for the proper normalization of the hypercharge generator of the
U(1)$_Y$ group.

For one SM Higgs doublet, $N=1$, there is no unique scale $\bar\mu$,
where
$\alpha_1^{-1}(\bar\mu)=\alpha_2^{-1}(\bar\mu)=\alpha_3^{-1}(\bar\mu)$.
However, if at some scale $\hat\mu$ new states start to make
additional contribution to the gauge coupling evolution, this
unification point can be found. Since the evolution of the gauge
coupling constants obeys linear behavior with $\ln\mu$, simple
formulas can be obtained
\begin{eqnarray}
\label{exact} \nonumber
  &&\hat{\mu} = M_Z\exp\left[-
  \frac{2\pi\,\epsilon_{ijk}(b_i-b_j)\alpha_k^{-1}(M_Z)}
  {\epsilon_{ijk}(b_i-b_j)\Delta b_k}\right],\\
  &&\bar\mu = M_Z\exp\left[-
  \frac{2\pi\,\epsilon_{ijk}(\Delta b_i-\Delta b_j)\alpha_k^{-1}(M_Z)}
  {\epsilon_{ijk}(b_i^0-b_j^0)\Delta b_k}\right], \\
\nonumber
  &&\alpha^{-1}(\bar\mu) =
  \frac{\epsilon_{ijk}\Delta b_i b_j^0\alpha_k^{-1}(M_Z)}
  {\sum_k\epsilon_{ijk}\Delta b_i b_j^0},
\end{eqnarray}
where $\Delta b_i=b_i-b_i^0$ are the differences between the
constants $b_i$ from eq.~(\ref{bi}) and their SM values $b_i^0$ at
$N=1$.

Since the differences $\Delta b_i$ are proportional to $N-1$ or
zero, it is obvious from eq.~(\ref{exact}) that the unification
scale $\bar\mu$ and the unified gauge coupling constant
$\alpha(\bar\mu)$ do not depend on $N$:
\begin{eqnarray}
\nonumber
  \bar\mu &=&
  M_Z\exp\frac{\pi[5\alpha_1^{-1}(M_Z)-3\alpha_2^{-1}(M_Z)
  -2\alpha_3^{-1}(M_Z)]}{22}\\
\nonumber
  &=& 5.09^{+0.09}_{-0.08}\times 10^{13} {\rm ~GeV}, \\
\nonumber
\alpha\!&^{-1}&(\bar\mu)=\frac{35\alpha_1^{-1}(M_Z)-21\alpha_2^{-1}(M_Z)
  +30\alpha_3^{-1}(M_Z)}{44}\\
  &=& 38.57\pm 0.03,
\end{eqnarray}
Although the solution, eq.~(\ref{exact}), always exists, the
physically acceptable result $\hat\mu > M_Z$ is possible only for
$N\geq8$. Therefore, the lightest states, which can provide
unification, correspond to $N=8$ and the scale
\begin{equation}\label{hatmu}
    \hat{\mu} = 692^{+144}_{-120} {\rm ~GeV}.
\end{equation}
In the following only this possibility will be discussed.

\section*{Weak-doublet spin-1 bosons}

\begin{figure}[h!]
\includegraphics[width=0.5\textwidth]{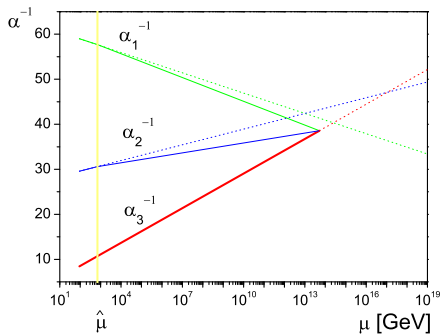}
\caption{\label{fig:1} One-loop evolution of the gauge coupling
constants in the SM (dotted lines) and with the extended $N=8$ Higgs
sector (solid lines).}
\end{figure}

Although the unification of the gauge coupling constants is reached
(Fig.~\ref{fig:1}), the extension of the SM with seven additional
Higgs doublets looks awkward. Here we will propose a different
interpretation of the given result.

In \cite{Dvali} it was shown that the introduction of the
weak-doublet spin-1 bosons $V_\mu=(Z^\star_\mu,W^{\star-}_\mu)$ with
the internal quantum numbers identical to the SM Higgs doublet is
motivated by the hierarchy problem. It means that each spin-0 Higgs
doublet is associated with the corresponding spin-1 doublet and vice
versa. Such spin-1 states were earlier introduced in \cite{doublets}
using the formalism of the chiral antisymmetric tensor fields.

Their interactions with SU(2)$_W\times$U(1)$_Y$ gauge fields are
similar to the interactions of the SM Higgs doublet due to identical
internal quantum numbers. The massive vector boson has three
physical degrees of freedom and contributes to the gauge coupling
evolution in the one-loop approximation three times more strongly
than the scalar boson. Therefore, introduction of one pair of scalar
and vector doublets is equivalent to the four Higgs doublets
content.

So, the solution with $N=8$ can be interpreted as an extension of
the SM Higgs sector with one additional Higgs doublet and two
corresponding vector doublets. That is exactly the set of fields
which was proposed in \cite{doublets}. It was shown~\cite{doublets}
that the second pair of scalar and vector doublets with opposite
hypercharges is necessary to cancel the chiral anomaly.

Due to their quantum numbers, in the leading order the vector
doublets can only have anomalous (magnetic moment type) interactions
with the SM fermions,
\begin{eqnarray}
\label{maincouplingdown} &&{1 \over M}  \,D_{\mu} V_{\nu}^{\rm c} \,
\left ( g^d_{LR} \, \overline{Q_L} \sigma^{\mu\nu} d_R \,
+ \, g^e_{LR} \, \bar{L}\sigma^{\mu\nu} e_R  \right )  \\
\label{maincouplingup} &+&{g^u_{LR} \over M}\,  D_{\mu} V_{\nu} \,
\overline{Q_L} \sigma^{\mu\nu} \, u_R \,+{\rm h.c.},
\end{eqnarray}
where $V_{\mu}^{\rm c} \, \equiv (-W_{\mu}^{*+}, \bar{Z}^{*}_{\mu})$
is the charge-conjugated doublet; $Q_L\equiv (u_L, d_L)$ and $L
\equiv (\nu_L, e_L) $ are the left-handed quark and lepton doublets
respectively. $D_{\mu}$ are the usual SU(2)$_W\times$U(1)$_Y$
covariant derivatives, and the obvious group and family indices are
suppressed. $M$ is the scale of new physics and $g^{u,d,e}_{LR}$ are
dimensionless constants.

The derivative couplings lead to previously unexplored angular
distributions \cite{collider} and to unique signatures for detection
of these bosons at the hadron colliders. Our project for their
search \cite{project} was accepted by the ATLAS Collaboration and
the corresponding analysis of the experimental data was performed.
In the simplest reference model \cite{refMod} used in the analysis,
the dimensionless constants were fixed to be proportional to the
electroweak gauge coupling and the {\em family universality} was
assumed. The scale of new physics was chosen to be equal to the mass
of the new bosons. The final Run-I ATLAS results \cite{ATLAS} put
the following 95\% CL limits on new boson masses:
\begin{equation}\label{limits}
    M_{W^\star} > 3.21~{\rm TeV},~~~~M_{Z^\star} > 2.85~{\rm TeV}.
\end{equation}

At first glance, these results exclude the possibility of existence
of the lightest states with masses $M\sim\hat\mu\approx 700$~GeV
(see eq.~(\ref{hatmu})). However, they were derived from analyzing
the final states with light leptons (electrons, muons and missing
neutrinos) as the clearest channels for new heavy resonance search
at hadron colliders. It was also assumed that the heavy resonance
had to be produced in direct fusion of the lightest quark-antiquark
pairs from the colliding protons. In other words, the quark-lepton
and {\em family universality} was assumed.

However, the assumption of the {\em family universality} is natural
for the vector fields from the adjoint representations of the gauge
group in order to avoid tree-level flavor-changing neutral currents,
whereas the scalar Higgs doublet from the fundamental representation
interacts mainly with the fermions from the third family, which is
the source of the flavor violation in Nature. Since the vector
doublets come along with the scalar doublets, it is more natural to
suggest a similar pattern of couplings for the vector doublets too.

The presence of two Higgs doublets can also be a source of
tree-level flavor-changing neutral currents. In order to prevent
that, we assume that one doublet pair couples only to up-type
quarks, while the other couples to down-type quarks and charged
leptons only \cite{GW}. Since the coupling strengths are
proportional to the mass of the coupled fermions, in this simple
model the $W^\star$ and $Z^\star$ bosons from the vector doublet
couple mainly to the right-handed $t$-quark singlet (see eq.
(\ref{maincouplingup}))
\begin{eqnarray}
\label{simpleW}
  && \frac{g^\star}{M_{W^\star}} \left (
  \partial_{\mu} W^{\star-}_{\nu} \, \overline{b_L}
  \sigma^{\mu\nu} t_R \,
  + \,  \partial_{\mu} W^{\star+}_{\nu} \,
  \overline{t_R} \sigma^{\mu\nu} b_L \right )\\
  +&& \frac{g^\star}{\sqrt{2}M_{Z^\star}} \left (
  \partial_{\mu} {\rm Re}Z^{\star}_{\nu}\, \bar{t}
  \sigma^{\mu\nu} t \,
  + \,  i\partial_{\mu} {\rm Im}Z^{\star}_{\nu} \,
  \bar{t} \sigma^{\mu\nu}\!\gamma^5 t \right ),
\label{simpleZ}
\end{eqnarray}
where Re$Z^\star$ and Im$Z^\star$ are the properly normalized
CP-even and CP-odd neutral states.

\section{Experimental signatures}

Theoretical and experimental aspects for extra scalar bosons search
from two-Higgs-doublet models are already extensively studied
\cite{twoHiggs}. Therefore, we concentrate here on a less-known
issue connected with the production and decay of vector doublets.

Interactions of vector doublets resemble scalar Higgs doublet
couplings. Therefore, experimental signatures should be within the
scope of those for the Higgs searches, although with obvious
differences due to different spins.

\begin{figure}[h!]
\includegraphics[width=0.23\textwidth]{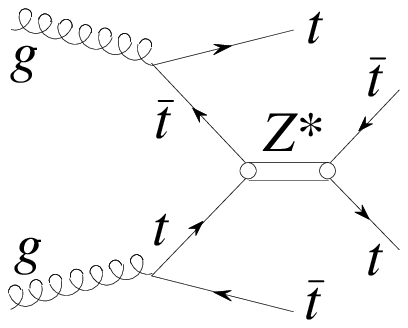}
\includegraphics[width=0.23\textwidth]{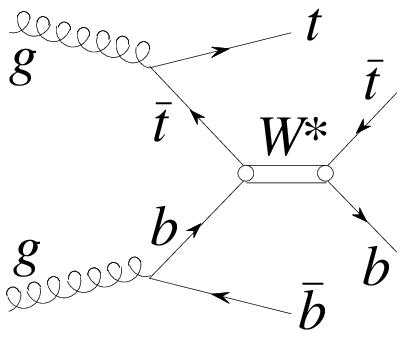}
\includegraphics[width=0.23\textwidth]{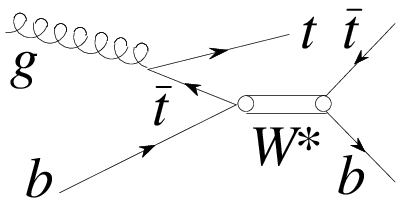}
\includegraphics[width=0.23\textwidth]{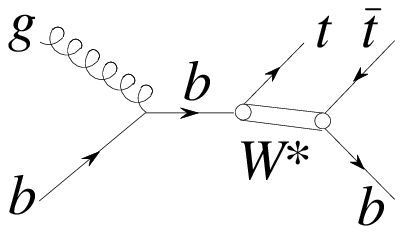}
\caption{\label{fig:2} Production of new vector bosons.}
\end{figure}

For example, the leading channel for the Higgs production at the
LHC, gluon-gluon fusion through the $t$-quark loop, is not operative
or suppressed for production of vector bosons due to the
Landau--Yang theorem~\cite{LY}. Vector fields cannot have nonzero
vacuum expectation value unless Lorentz symmetry violation exists.
Therefore, analogs of the Higgs-strahlung and weak vector boson
fusion production processes are also absent for the vector doublet
boson production. For the same reason the new vector boson cannot
decay into two photons or two $Z$ bosons, which are used as very
clean channels for precise reconstruction of the Higgs mass. The
only highly suppressed processes of heavy quark--antiquark fusion
can produce resonantly the new vector bosons (Fig.~\ref{fig:2}).

The final signature of the first process (the upper left panel of
Fig.~\ref{fig:2}) $gg\to t\bar{t}Z^\star\to t\bar{t}t\bar{t}$ is
already in the sights of the standard model group~\cite{CMS4t} and
the exotic group~\cite{23leptons}. Although the SM Higgs boson mass
is below the threshold for the $t\bar{t}$-quarks production, the
process for the associated Higgs production with a top-quark pair in
multi-lepton final states
\cite{ttHmultilepton} can mimic the signature of the first process.

Since the SM cross section $\sigma^{\rm
SM}_{t\bar{t}t\bar{t}}\approx1$~fb at $\sqrt{s}=8$~TeV is very
small~\cite{SM4t} we can put upper limit on the coupling constant of
the neutral vector bosons from eq.~(\ref{simpleZ}) $g^\star < 2$
assuming $M_{Z^\star}\approx 700$~GeV from the direct 95\% CL
constraint $\sigma^{\exp}_{t\bar{t}t\bar{t}}<32$~fb~\cite{CMS4t}.
From here on the CalcHEP package~\cite{CalcHEP} is used for all
numerical estimations.

The final signature of the second process (the upper right panel of
Fig.~\ref{fig:2}) $gg\to t\bar{b}W^{\star-}\to t\bar{t}b\bar{b}$
coincides with the associated production of the Higgs boson with a
top-quark pair and its decay into bottom quarks, although with
absolutely different kinematics. Both collaborations have already
searched for this final state~\cite{ttbb}.

The third process for the $W^\star$ production (the bottom panel of
Fig.~\ref{fig:2}) $gb\to t W^{\star-}\to tb\bar{t}$ has a cross
section comparable with the second process and leads to a yet
unexplored final signature.\footnote{Only recently the CMS
Collaboration has announced their analysis~\cite{CMSttb} using this
final state for a charged Higgs boson search.} This signature is
much simpler than the previous ones and can be used even for direct
reconstruction of the $b\bar{t}$-invariant mass.

\begin{figure}[h!]
\includegraphics[width=0.47\textwidth]{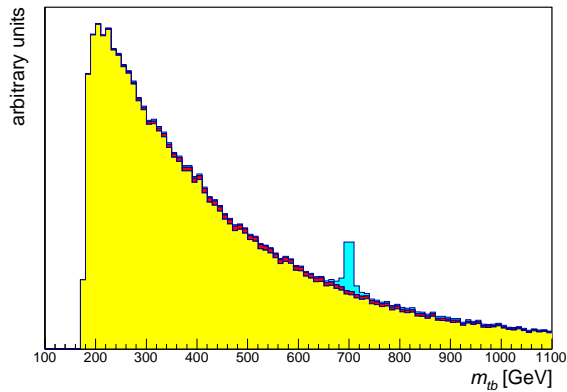}
\caption{\label{fig:3}  The $tb$ and $b\bar{t}$-invariant mass
distributions at $\sqrt{s}=13$~TeV.}
\end{figure}

As far as it is impossible to disentangle jets produced by quarks or
antiquarks, we will use both possibilities, $tb$ and $b\bar{t}$,  in
order to construct the invariant mass, $m_{tb}$. The corresponding
histograms at $\sqrt{s}=13$~TeV are shown in Fig.~\ref{fig:3}.
The SM contribution is represented by a monotonically decreasing
histogram (light yellow color). The nonresonant $tb$ contribution
(dark red color) is also distributed over the whole region of
possible invariant masses and is negligible under the peak from the
$b\bar{t}$ contribution (light cyan color).

\section{Discussions}

In this paper we have shown that the extension of the SM Higgs
sector leads to a unification scale around $5\times10^{13}$ GeV.
This value has many specific features.

For example, if the Majorana mass of a sterile right-handed neutrino
is of the order of the unification scale, then the light neutrino
states should have the mass of the expected order $m_\nu\sim
v^2/2\bar\mu\approx 0.6$~eV due to the see-saw
mechanism~\cite{seesaw}. Here $v$ is the vacuum expectation value of
the Higgs field. This not so high unification scale is closer to the
allowed heavy Majorana neutrino masses for successful baryogenesis
through leptogenesis~\cite{leptogen}. This scale does not destroy
naturality from the Planck scale~\cite{agravity} $\delta m_h\sim
\bar\mu^3/(4\pi)^3M^2_{\rm Pl}\approx 0.5$~GeV.

On the other hand, the new lightest states at the scale
$\hat\mu\approx 700$~GeV maintain naturality, solving the hierarchy
problem~\cite{Dvali}. The introduction of the spin-1 doublets with
the vector degrees of freedom replaces the introduction of many
scalar states with degenerate masses. According to \cite{chile}, it
is an important feature to avoid contradiction with precision
low-energy data.

\section{Conclusions}

It is shown that the gauge coupling unification can be achieved
in the one-loop approximation by extension of the Standard Model
scalar sector only with extra Higgs doublets. As a result, the
unification scale is lower than other scales known in the literature
and does not depend on the extra Higgs doublets, whereas a new
physics scale, at which new scalar degrees of freedom become active,
can be well below 1~TeV. However, this scale of new physics is
reached at total number of 8 Higgs doublets, which looks very
awkward.

Therefore, we assume that spin-1 vector bosons can play the role of
some of scalar degrees of freedom. In this case we get a compact
fields content: two Higgs and two spin-1 doublets. However, such
light states were not found in the first LHC run. The reason, as we
see it, is in accepting the hypothesis of family universality of
vector doublet interactions with quarks and leptons. If the vector
doublet interactions resemble the Higgs fermion couplings, the new
spin-1 bosons cannot be produced in light quark--antiquark
annihilation from the proton beams and cannot decay into light
lepton pairs as well.

This means that the production and the decay of the new heavy bosons
should be associated only with heavy $b$- and $t$-quarks. Moreover,
the increasing gluon luminosity due to higher centre-of-mass
energies in the second LHC run will lead to an order of magnitude
higher cross sections for the considered processes than in the first
LHC run. In conclusion, we would like to stress out that the new
channel $gb\to t W^{\star-}\to tb\bar{t}$ can be very useful for
early new physics search.

\end{document}